\documentclass[12pt]{article}
\pdfoutput=1

\usepackage{graphicx}
\usepackage{color,amssymb,amsmath}
\usepackage{slashed}
\usepackage[a4paper, total={16.5cm,22cm}]{geometry}
\usepackage{hyperref}
\usepackage{booktabs}
\usepackage{braket}
\usepackage{slashed}
\usepackage[separate-uncertainty=true]{siunitx}
\usepackage{cite}
\usepackage{cancel}
\usepackage[utf8]{inputenc} 

\usepackage[compat=1.1.0]{tikz-feynman}
\tikzfeynmanset{
every edge={thick},
}

\hypersetup{
  colorlinks = true,
  linkcolor = blue,
  citecolor = blue,
}

\newcommand\Order{\mathop{\mathcal{O}}}
\newcommand\TO{\text{--}}
\newcommand\unit[1]{\,\mathrm{#1}}
\newcommand\GeV{\unit{GeV}}
\newcommand\TeV{\unit{TeV}}
\newcommand\fb{\unit{fb}}

\newcommand\ifb{\unit{fb^{-1}}}

\let\tilde\widetilde

\begin{document}

\begin{titlepage}
\begin{flushright}
UT--17--14\\
IPMU17--0061\\
KEK--TH--1973
\end{flushright}
\vskip 1.5cm
\begin{center}
  {\Large \bf Probing minimal SUSY scenarios\\
    in the light of muon $g-2$ and dark matter\\
  }
\vskip 1.5cm
{
  Motoi Endo,$^{(a,b,c)}$
  Koichi Hamaguchi$^{(d,c)}$,
  Sho Iwamoto$^{(e)}$,
  and
  Keisuke Yanagi$^{(d)}$
}
\vskip 0.9cm
{\it $^{(a)}$ KEK Theory Center, IPNS, KEK, Tsukuba, Ibaraki 305--0801, Japan
\vspace{0.2cm}
\par
$^{(b)}$ The Graduate University of Advanced Studies (Sokendai),\\
Tsukuba, Ibaraki 305--0801, Japan
\vspace{0.2cm}
\par
$^{(c)}$ Kavli Institute for the Physics and Mathematics of the Universe (Kavli IPMU), \\
University of Tokyo, Kashiwa 277--8583, Japan
\vspace{0.2cm}
\par
$^{(d)}$ Department of Physics, University of Tokyo, Bunkyo-ku, Tokyo 113--0033, Japan
\vspace{0.2cm}
\par
$^{(e)}$ Physics Department, Technion---Israel Institute of Technology, Haifa 3200003, Israel
}
\vskip 2.5cm
 \abstract{
We study supersymmetric (SUSY) models in which the muon $g-2$ discrepancy and the dark matter relic abundance are simultaneously explained.
The muon $g-2$ discrepancy, or a 3$\sigma$ deviation between the experimental and theoretical results of the muon anomalous magnetic moment, can be resolved by SUSY models, which implies at least three SUSY multiplets have masses of $\Order(100)\GeV$.
In particular, models with the bino, higgsino and slepton having $\Order(100)\GeV$ masses are not only capable to explain the muon $g-2$ discrepancy but naturally contains the neutralino dark matter with the observed relic abundance.
We study constraints and future prospects of such models; in particular, we find that the LHC search for events with two hadronic taus and missing transverse momentum can probe this scenario through chargino/neutralino production.
It is shown that almost all the parameter space of the scenario can be probed at the high-luminosity LHC, and a large part can also be tested at the XENON1T experiment as well as at the ILC.

 }

\end{center}
\end{titlepage}

\section{Introduction}
\label{sec:introduction}

Supersymmetry (SUSY) is one of the most attractive candidates of physics
beyond the Standard Model (SM).
The electroweak scale is stabilized against the radiative corrections,
the gauge coupling unification becomes much better than in the SM,
and the lightest SUSY particle (LSP) can be the dominant component of the dark matter (DM).
On the other hand, the stringent constraints from the LHC searches as well as the 125 GeV Higgs boson mass may imply that the SUSY particles, in particular the colored ones, are much heavier than $\Order(0.1\TO1)\TeV$.

However, there are still motivations to consider SUSY with $\Order(100)\GeV$ sparticles: (i) the naturalness, (ii) the neutralino DM, and, although not conclusive, (iii) the discrepancy of the anomalous magnetic moment ($g-2$) of the muon.
Among these motivations, the naturalness requires at least higgsino, stop, and gluino to be light, and such a mass spectrum is severely constrained by the recent SUSY search results (see, e.g., Ref.~\cite{Buckley:2016kvr} for a recent study).
We focus on the other two motivations, (ii) the neutralino DM and (iii) the muon $g-2$ discrepancy, and investigate minimal SUSY models that can explain these two simultaneously.\footnote{%
For recent studies of prospects for SUSY models in light of DM and the muon $g-2$, 
see, e.g., Refs.~\cite{Choudhury:2017fuu, Kobakhidze:2016mdx,Chakraborti:2015mra, Ajaib:2015yma, Padley:2015uma,
 Wang:2015rli, Kowalska:2015zja, Li:2014dna, Das:2014kwa, Chakraborti:2014gea, Cheng:2013hna}.
}

The value of the muon $g-2$ is reported by the Brookhaven E821 experiment as~\cite{Bennett:2006fi,Roberts:2010cj}
\begin{equation}
   a_\mu(\text{exp}) = \num{11659208.9\pm6.3e-10}\,,
\end{equation}
where $a_\mu = (g_\mu-2)/2$.
It is compared with the SM prediction
\begin{equation}
  a_\mu(\text{SM}) =
  \begin{cases}
  \num{11659182.8\pm4.9e-10}\ \text{\cite{Hagiwara:2011af}}, \\
  \num{11659180.2\pm4.9e-10}\ \text{\cite{Davier:2010nc}},
  \end{cases}
\end{equation}
where $a_\mu(\text{LbL}) = \num{10.5\pm2.6e-10}$ is used for the estimation of the hadronic light-by-light contributions \cite{Prades:2009tw}.
The difference is evaluated as
$a_\mu(\text{exp}) - a_\mu(\text{SM}) = \num{26.1\pm8.0e-10}$~\cite{Hagiwara:2011af}
or $\num{28.7\pm8.0e-10}$~\cite{Davier:2010nc}.\footnote{%
  If we use another result $a_\mu(\text{LbL}) = \num{11.6\pm4.0e-10}$ calculated in Ref.~\cite{Jegerlehner:2009ry},
  the difference becomes
  $\num{25.0\pm8.6e-10}$ or $\num{27.6\pm8.6e-10}$ with the contributions of hadronic vacuum polarization $a_\mu^{\text{had-VP}}$ calculated in Refs.~\cite{Hagiwara:2011af} and \cite{Davier:2010nc}, respectively.}
Therefore, the experimental result is larger than the SM expectation by the $> 3\sigma$ level. Such a large contribution can be explained in SUSY models with $\Order(0.1\TO1)\TeV$ smuons and charginos/neutralinos~\cite{Lopez:1993vi,Chattopadhyay:1995ae,Moroi:1995yh}.

In order to explain the muon $g-2$ discrepancy, at least three SUSY multiplets must be as light as $\Order(100)\GeV$.
In this letter, we consider minimal scenarios, i.e., models in which only three SUSY multiplets have $\Order(100)\GeV$ masses and the other SUSY particles are much heavier.
There are four minimal scenarios, which are characterized by the light SUSY multiplets:
\begin{description}
  \setlength{\leftskip}{2cm}
\item[BHR]
  bino, higgsino, and right-handed slepton,
\item[BHL]
  bino, higgsino, and left-handed slepton,
\item[BLR]
  bino, left- and right-handed sleptons,
\item[WHL]
  wino, higgsino, and left-handed slepton.
\end{description}
In each scenario, the three light SUSY multiplets yield loop-level contributions to the muon $g-2$ to resolve the discrepancy, as we will see in Sec.~\ref{sec:setup}.

The WHL scenario, however, cannot explain the DM abundance because the thermal wino (higgsino) DM predicts its mass around 2.9 TeV \cite{Hisano:2006nn} (1 TeV \cite{Cirelli:2007xd}), which is too large for the muon $g-2$.
The phenomenology of the BLR scenario has been investigated comprehensively in Ref.~\cite{Endo:2013lva}, although the DM physics was not discussed in detail.

In this letter, we study the BHR and BHL scenarios in the light of the muon $g-2$ discrepancy and the DM abundance.
The DM is the lightest neutralino $\tilde N_1$ dominated by the bino component, whose mass as well as those of the higgsino and the sleptons are $\Order(100)\GeV$.
The scenarios can be tested at the LHC, by DM direct detections, and at the ILC.
In particular, we find that the LHC search for events with two hadronic taus and missing transverse momentum, whose original target is the direct stau production, can  probe this scenario through production of chargino/neutralino decaying into taus.
It is shown that almost all the parameter space of the scenario can be probed by the high-luminosity LHC (HL-LHC), and a large part can also be tested by the XENON1T experiment as well as the ILC.
It is emphasized that, as the SUSY particles that are irrelevant to the muon $g-2$ and DM are assumed to be heavy and decoupled, our conclusion is quite general and independent of details of the models.


\section{Setup}
\label{sec:setup}

The following SUSY parameters are responsible for the muon $g-2$:
\begin{equation}
  \label{eq:simp-params}
  M_1, M_2, \mu, m_L^2, m_R^2, \tan\beta,
\end{equation}
where $M_1$, $M_2$, $m_L$ and $m_R$ are the soft masses for the bino, wino, left-handed and right-handed sleptons, respectively. $\mu$ is the higgsino mass parameter, and $\tan\beta=\langle H_u\rangle/\langle H_d\rangle$ is the ratio of the vacuum expectation values of the up- and down-type Higgs.
For simplicity, we take 
the slepton soft masses to be universal and flavor-blind, and assume that the effects of the $A$-terms are negligible.\footnote{The case of non-universal slepton masses is briefly discussed in Sec.~\ref{sec:summary-discussion}.} 
We further assume that the complex phases of soft parameters are neglected,
and take $M_1$ and $M_2$ to be positive, while $\mu$ can have either sign, following the convention of Ref.~\cite{Martin:1997ns}.
To obtain sizable contributions to the muon $g-2$,
we take $\tan\beta = 40$ throughout our analysis.
All the analyses are done with tree-level masses and couplings,
including the $D$-term contributions to the slepton masses.

Considering the one-loop diagrams with gauge eigenstates, the SUSY contributions to the muon $g-2$ are classified into four types: BHR, BHL, BLR, and WHL, and for large $\tan\beta$ the contributions are respectively approximated as~\cite{Moroi:1995yh}\footnote{In the numerical calculation, we do not use these approximation but the full one-loop level formula in terms of the mass eigenstates. For two-loop SUSY contributions, see Refs.~\cite{Fargnoli:2013zda,Fargnoli:2013zia,Athron:2015rva}.}
\begin{align}
  a_{\mu}(\mathrm{BHR})
  &\simeq - \frac{\alpha_Y}{4\pi} \frac{m_{\mu}^2}{M_1\mu} \tan\beta\cdot
    f_N\left(\frac{M_1^2}{m_{\tilde{\mu}_R}^2}, \frac{\mu^2}{m_{\tilde{\mu}_R}^2} \right),
    \label{eq:BHR} \\
  a_{\mu}(\mathrm{BHL})
  &\simeq \frac{\alpha_Y}{8\pi} \frac{m_{\mu}^2}{M_1\mu} \tan\beta\cdot
    f_N\left(\frac{M_1^2}{m_{\tilde{\mu}_L}^2}, \frac{\mu^2}{m_{\tilde{\mu}_L}^2} \right),
    \label{eq:BHL}\\
  a_{\mu}(\mathrm{BLR})
  &\simeq \frac{\alpha_Y}{4\pi} \frac{m_{\mu}^2M_1\mu}{m_{\tilde{\mu}_L}^2m_{\tilde{\mu}_R}^2}
    \tan\beta\cdot
    f_N\left(\frac{m_{\tilde{\mu}_R}^2}{M_1^2}, \frac{m_{\tilde{\mu}_R}^2}{M_1^2} \right),
    \label{eq:BLR} \\
  a_{\mu}(\mathrm{WHL1})
  &\simeq -\frac{\alpha_2}{8\pi} \frac{m_{\mu}^2}{M_2\mu} \tan\beta\cdot
    f_N\left(\frac{M_2^2}{m_{\tilde{\mu}_L}^2}, \frac{\mu^2}{m_{\tilde{\mu}_L}^2} \right),
    \label{eq:WHL1}\\
  a_{\mu}(\mathrm{WHL2})
  &\simeq \frac{\alpha_2}{4\pi} \frac{m_{\mu}^2}{M_2\mu} \tan\beta\cdot
    f_C\left(\frac{M_2^2}{m_{\tilde{\nu}_{\mu}}^2}, \frac{\mu^2}{m_{\tilde{\nu}_{\mu}}^2} \right),
    \label{eq:WHL2}                              
\end{align}
where the loop functions are given by
\begin{align}
    \label{eq:loop-aprox}
    f_C(x,y)
    &=xy\left[
      \frac{5-3(x+y)+xy}{(x-1)^2(y-1)^2} - \frac{2\ln x}{(x-y)(x-1)^3}+\frac{2\ln y}{(x-y)(y-1)^3}
      \right],
    \\
    f_N(x,y)
    &= xy\left[
      \frac{-3+x+y+xy}{(x-1)^2(y-1)^2} + \frac{2x\ln x}{(x-y)(x-1)^3}-\frac{2y\ln y}{(x-y)(y-1)^3}
      \right].
\end{align}

As discussed in the Introduction, we consider minimal scenarios, where only three SUSY multiplets have $\Order(100)\GeV$ masses and the other SUSY particles are much heavier.
Then, the simultaneous explanation of the muon $g-2$ discrepancy and the DM abundance under the thermal relic density requires that the LSP should be bino-like with the mass of $\Order(100)\GeV$.
In this letter, we study in detail the following two scenarios (bino-higgsino-slepton scenarios).
Other possibilities are briefly discussed in Sec.~\ref{sec:summary-discussion}.

\paragraph{BHR scenario}
The bino, the higgsino, and the right-handed sleptons are as light as $\Order(100)\GeV$ and
$M_1$, $\mu$, and $m_R$ are the relevant parameters.
The other soft masses are taken to be decoupled.
Since $a_{\mu}(\mathrm{exp}) - a_{\mu}(\mathrm{SM})$ is positive,
we need negative $\mu$ to obtain the positive contribution (see Eq.~\eqref{eq:BHR}).
In the present analysis, we focus on the parameter region of $M_1\ge 100$~GeV.
There are two possibilities of DM in this scenario:
bino-slepton coannihilation, and the well-tempered bino/higgsino~\cite{ArkaniHamed:2006mb}.
The well-tempered neutralino, however, is disfavored because it results in a large DM-nucleus scattering cross section that is inconsistent with DM direct detections.\footnote{The blind-spot suppression for the direct detection~\cite{Cheung:2012qy} does not work, since we consider large $\tan\beta$ to enhance the muon $g-2$.}
We thus focus on the bino DM through bino-slepton coannihilation.
Based on these, we adopt the following procedure to constrain the parameter space:
at each point on the $(M_1,\mu)$ plain, we find $m_R$ that provides the correct DM relic abundance,
$\Omega_{\mathrm{LSP}}=\Omega_{\mathrm{DM}}$.
Then  we calculate the contribution to the muon $g-2$ and experimental bounds and prospects from the DM direct detection, LHC, and ILC.

\paragraph{BHL scenario}
The bino, the higgsino, and the left-handed sleptons are the light particles and the relevant parameters are $M_1$, $\mu$, and $m_L$.
As in the BHR scenario, the other soft masses are taken to be decoupled.
To obtain the positive $g-2$ contribution from Eq.~\eqref{eq:BHL},
we need positive $\mu$.
Since the constraint from DM direct detection is determined mainly by the higgsino component of the LSP,
the situation is similar to the BHR scenario, and we consider only the bino-slepton coannihilation for DM.
The same procedure as in the BHR scenario is thus adopted:
we first find $m_L$ that provides the correct relic abundance,
and then study the experimental bounds on the $(M_1,\mu)$ plain.
We note that, as we will discuss in Sec.~\ref{subsec:M2}, the wino soft mass $M_2$ is important for the muon $g-2$
in this scenario even when $M_2>1\TeV$,
since the wino contribution, i.e., the sum of Eqs.~\eqref{eq:WHL1} and \eqref{eq:WHL2}, is positive and still sizable in presence of light left-handed slepton.

\section{Prospects for the bino-higgsino-slepton scenarios}
\label{sec:main}

\begin{figure}[t]
  \centering
  \begin{minipage}{0.5\linewidth}
    \includegraphics[width=1.0\textwidth]{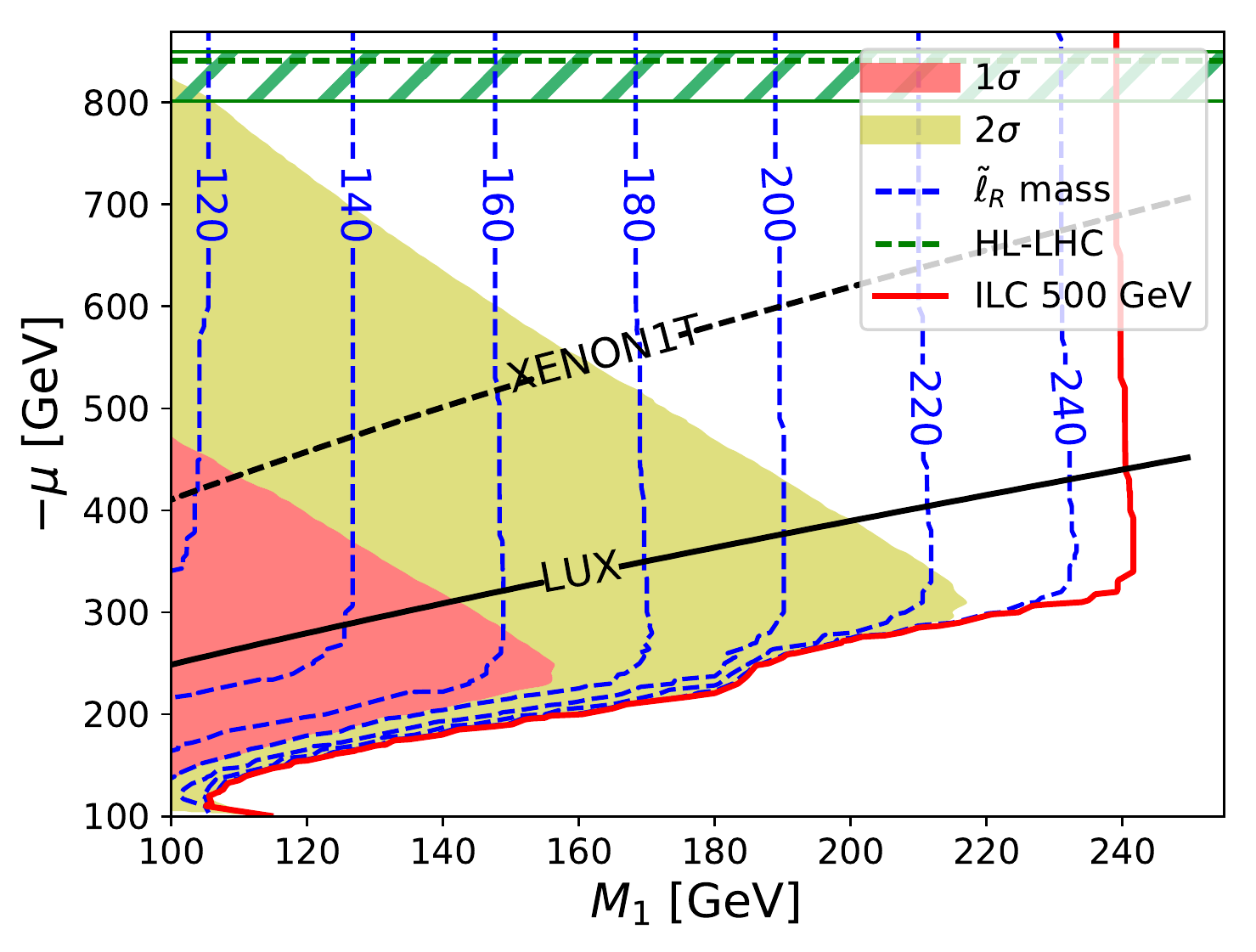}
  \end{minipage}%
  \begin{minipage}{0.5\linewidth}
    \includegraphics[width=1.0\textwidth]{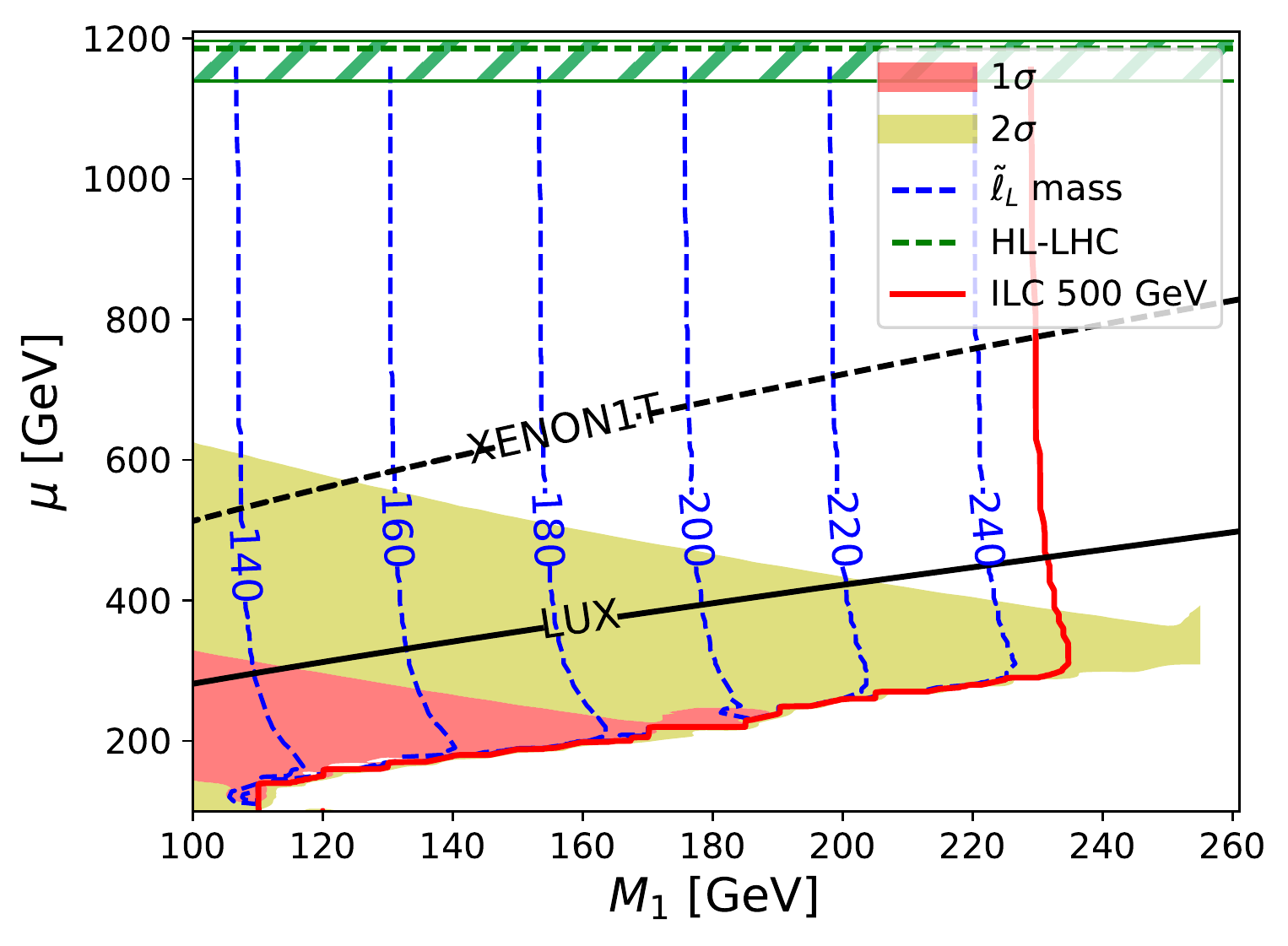}
  \end{minipage}
 \caption{The parameter region of our interest for the BHR scenario (left panel) and BHL scenario (right panel), together with experimental constraints and future prospects.
    We use $\tan\beta=40$ and $M_2=\SI{3}{TeV}$.
    The blue contours show the slepton mass
    $m_{\tilde{\ell}_R}$ or $m_{\tilde{\ell}_L}$ that gives $\Omega_{\mathrm{LSP}}=\Omega_{\mathrm{DM}}$.
    With the slepton mass, the muon $g-2$ discrepancy is explained within $1\sigma$ ($2\sigma$) uncertainty in the red (yellow) regions.
    The regions below the solid (dashed) lines are excluded (will be probed) by the LUX (XENON1T) experiment with 90\% confidence level.
    The regions below the green dashed lines will be probed
    by the HL-LHC with $\sqrt s=14\TeV$ and $\int\mathcal L=3000\ifb$, assuming 30\% systematic uncertainty from SM background;
    the green hatched regions correspond to different systematic uncertainties between 20\% and 50\%.
    The red solid line corresponds to $m_{\tilde{\ell}}=248\GeV$, which will be probed at the ILC with $\sqrt s=500\GeV$.
 }
  \label{fig:result}
\end{figure}

In this section, we study experimental constraints and future sensitivities to the BHR and BHL scenarios.
The results for the BHR (BHL) scenario are summarized in the left (right) panel of Fig.~\ref{fig:result}; the parameter regions that explain the muon $g-2$ discrepancy together with the DM relic abundance, the constraint from the latest LUX result, and the future prospects of XENON1T, HL-LHC, and ILC experiments are shown. We will explain each of the constraints and prospects in the following subsections. We also briefly discuss the impact of wino mass $M_2$ in Sec.~\ref{subsec:M2}.

\subsection{DM abundance and muon $g-2$}
The relic abundance of the bino-like DM is calculated with \texttt{micrOMEGAs 4.3.2} \cite{Barducci:2016pcb}.
For each model point on the $(M_1, \mu)$ plain, the slepton mass parameter $m_R$ or $m_L$ is tuned so that the bino-slepton coannihilation provides the correct relic abundance.
We fix $M_2 = \SI{3}{TeV}$, $\tan\beta=40$, and the other soft masses to be 5 TeV.
The resulting slepton (selectron/smuon) mass,
$m_{\tilde{\ell}_R}$ or $m_{\tilde{\ell}_L}$, is shown by blue contours in Fig.~\ref{fig:result}.\footnote{%
 We use $\ell$ to denote the first- and second-generation leptons ($e$ and $\mu$), while $l$ is used for $e$, $\mu$, and $\tau$.}
Typical mass difference between the LSP and the next-to-lightest SUSY particle (NLSP) is 10 GeV.
Then, we calculate the muon $g-2$ contribution using the determined slepton masses.
The red (yellow) regions in Fig.~\ref{fig:result} explain the muon $g-2$ discrepancy within $1\sigma$ ($2\sigma$) uncertainty.

\subsection{DM spin-independent scattering}
\label{sec:dm-direct-detection}

Since we assume hierarchical spectra in which the squarks and gluino are too heavy to affect the cross section on nucleus,
the spin-independent cross section is dominated by the tree-level neutralino-quark interaction via the SM Higgs exchange.
The quark-Higgs interaction is with Yukawa coupling,
while neutralino-Higgs interaction is determined by the bino-higgsino mixing.
The interaction Lagrangian between the neutralino DM and the Higgs and between quarks and Higgs are respectively given by
\begin{align}
   \label{eq:higgs-int}
  \mathcal{L}_{h\tilde{N}_1\tilde{N}_1} &= \frac{1}{2}\lambda^h h \overline{\tilde{N}_1} \tilde{N}_1,&
  \mathcal{L}_{hqq} &= \frac{g_2m_q}{2m_W} h\bar{q}q,
\end{align}
where the neutralino-Higgs coupling $\lambda^h$ is approximated as\footnote{In the numerical calculation, we use the exact formula at tree-level.}
\begin{equation}
  \label{eq:n-h-coupl}
  \lambda^h \simeq g_1\left(\frac{\mu\sin2\beta + M_1}{\mu^2-M_1^2}m_Zs_W
    +\Order\left(\frac{m_Zs_W}{\mu}\right)^2 \right).
\end{equation}
In the limit of negligible momentum transfer,
the effective Lagrangian between the neutralino DM and the proton or neutron is written as
$\mathcal{L} = f_p\overline{\tilde{N}_1} \tilde{N}_1\bar{p}p + f_n\overline{\tilde{N}_1} \tilde{N}_1 \bar{n}n$,
where \cite{Shifman:1978zn}
\begin{equation}
  \label{eq:fp}
   f_{p(n)}
   = m_{p(n)}\cdot \frac{\lambda^h}{2m_h^2}\cdot \frac{g_2}{2m_W}
   \left(
  \frac{2}{9} + \frac{7}{9}\sum_{q=u,d,s} f_{T_q}^{p(n)}
  \right).
\end{equation}
For the nucleon mass fraction $f_{T_q}^{p(n)}$,
we use the default values of \texttt{micrOMEGAs 4.3.2} listed in Table~\ref{tab:fp-fn-num}.
\begin{table}
  \centering
  \begin{tabular}{ccc}
    \toprule
    & $f_{T_q}^p$ & $f_{T_q}^n$ \\
    \midrule
   $u$ & 0.0153 & 0.0110 \\
    $d$ & 0.0191 & 0.0273 \\
    $s$ & 0.0447 & 0.0447 \\
    \bottomrule
  \end{tabular}
  \caption{Nucleon form factors used in our calculation of the spin-independent cross section, which are the default values of \texttt{micrOMEGAs 4.3.2} \cite{Barducci:2016pcb}.}
  \label{tab:fp-fn-num}
\end{table}
The spin-independent cross section per proton (neutron) is given by
\begin{equation}
  \label{eq:si-xect}
  \sigma_{p(n)} = \frac{4}{\pi}f_{p(n)}^2m_{p(n)}^2 \left(1+\frac{m_{p(n)}}{m_{\tilde{N}_1}} \right)^{-2},
\end{equation}

In Fig.~\ref{fig:result},
the current constraint from the LUX experiment \cite{Akerib:2016vxi} is shown by the black solid lines, below which the model is excluded with 90\% confidence level (CL). The future sensitivity of the XENON1T experiment~\cite{Aprile:2015uzo} is also shown by the black dashed lines.
We see that the $1\sigma$ parameter region of the BHL scenario is mostly excluded by the LUX experiment, and XENON1T will probe most of the $2\sigma$ region.
Meanwhile, the model points of the BHR scenario are still widely allowed, and the XENON1T experiment will probe most of the $1\sigma$ region.

\subsection{LHC search}
\label{sec:lhc-search}

\subsubsection{Recasting the stau search}

Let us now consider LHC searches for neutralinos, charginos and sleptons.
As we focus on the bino-slepton coannihilation scenario, the LSP and sleptons, being the NLSP, are very degenerate with a typical mass difference of $10\GeV$.
Therefore, the process
\begin{align}
  \label{eq:slep-prod}
  pp &\to \tilde{l}\tilde{l} \to l\tilde{N}_1l\tilde{N}_1,
\end{align}
produces soft leptons only, which makes searches for the direct production of sleptons difficult.
Here, $\tilde{l}$ ($l$) denotes sleptons (leptons) including stau (tau), $\tilde{N}_{i}$ is the $i$-th lightest neutralino, and $\tilde{C}_1^\pm$ is the lighter chargino.

\begin{figure}
  \centering
  \begin{tabular}[t]{cc}
    \begin{minipage}{0.5\linewidth}
    \includegraphics[width=1.0\textwidth]{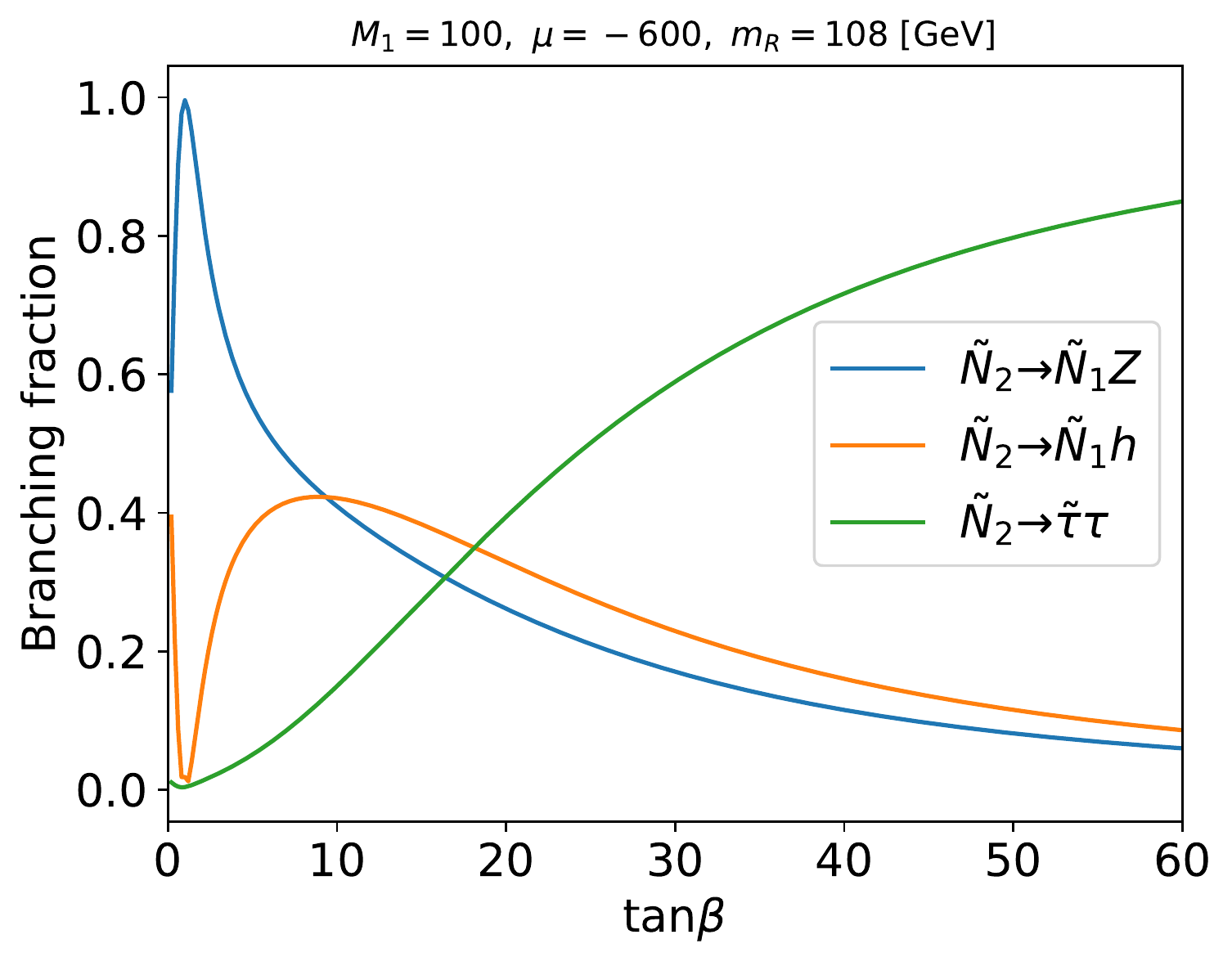}
  \end{minipage} &
   \begin{minipage}{0.5\linewidth}
    \includegraphics[width=1.0\textwidth]{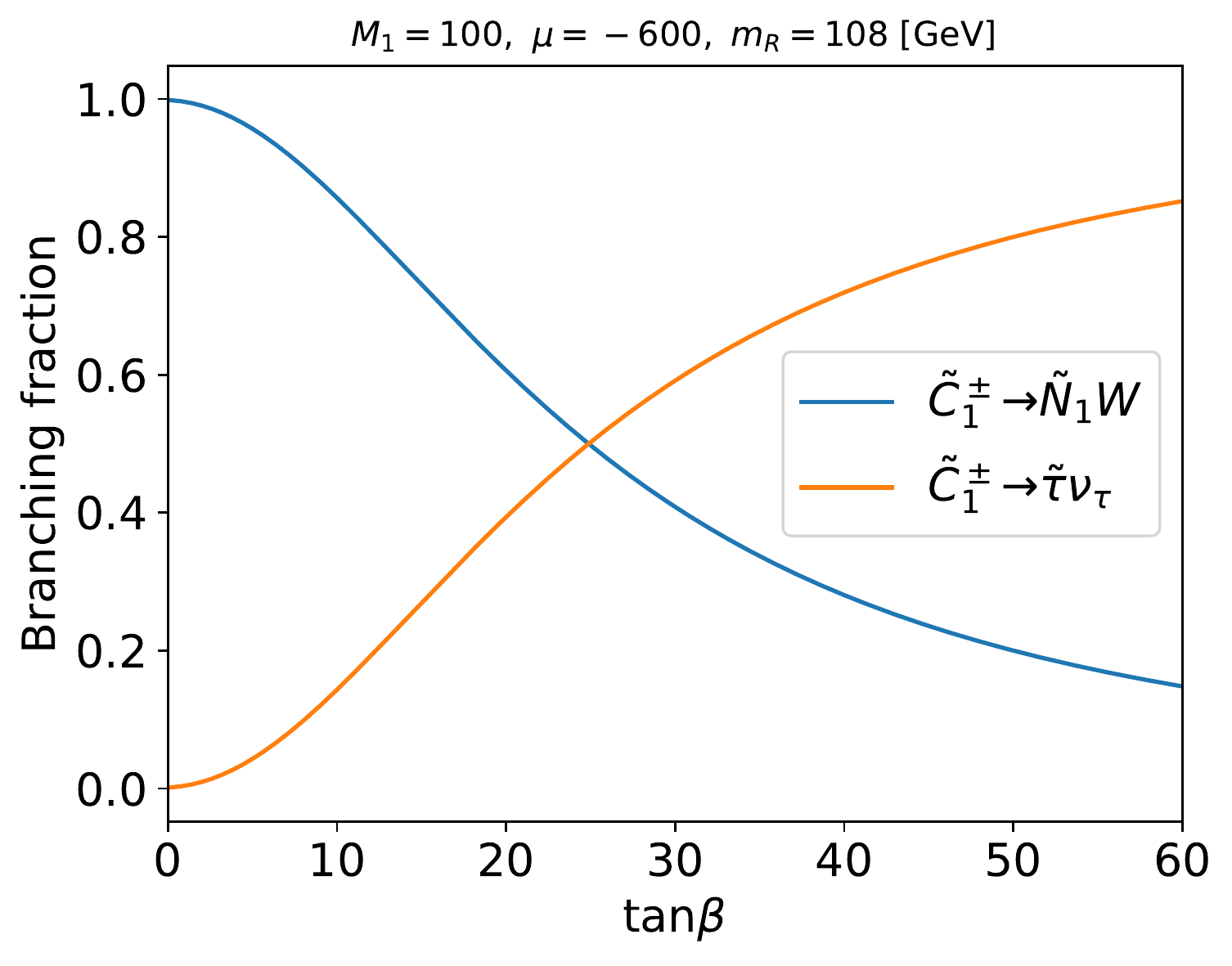}
  \end{minipage}\\
    \begin{minipage}{0.5\linewidth}
    \includegraphics[width=1.0\textwidth]{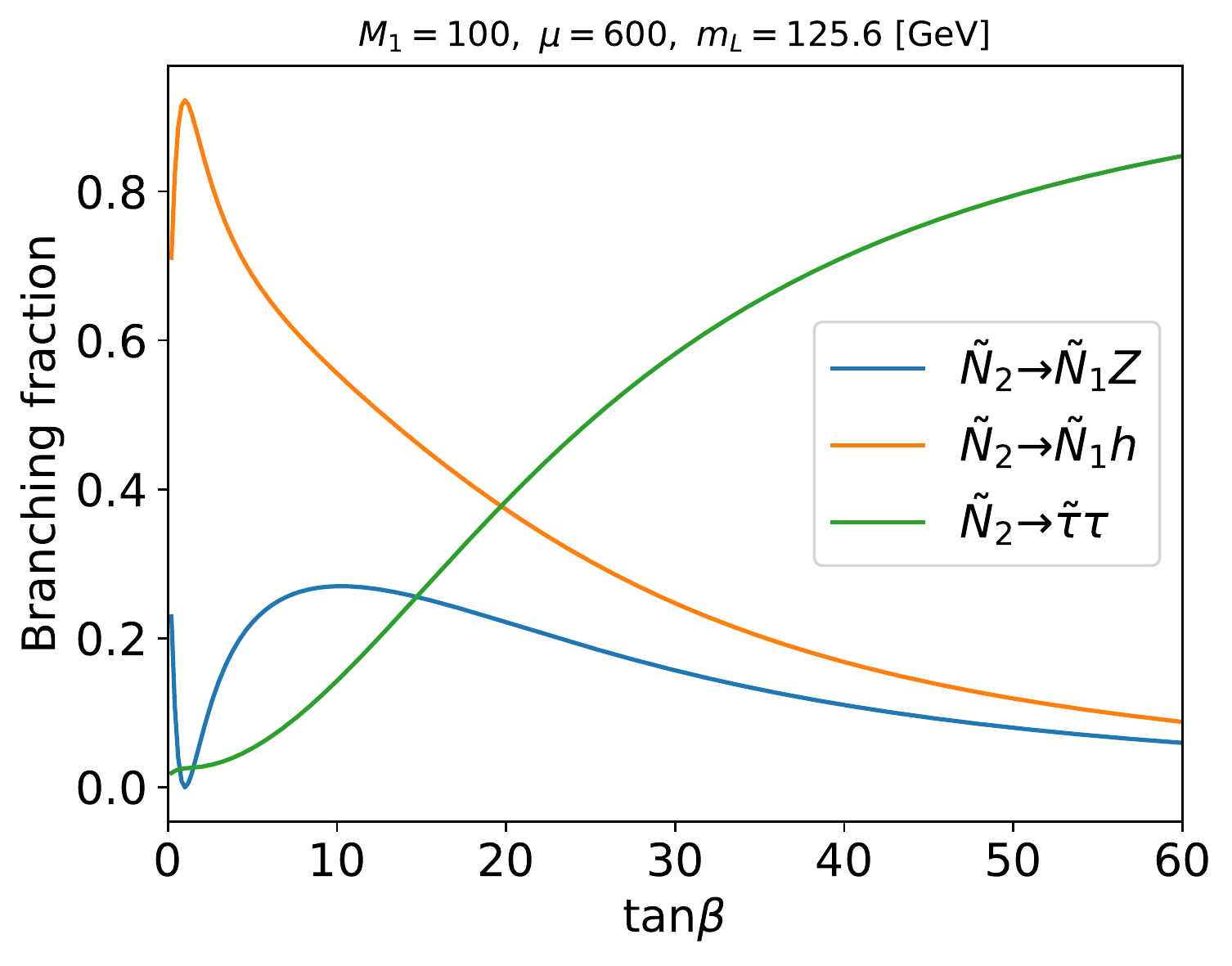}
  \end{minipage}  &
  \begin{minipage}{0.5\linewidth}
    \includegraphics[width=1.0\textwidth]{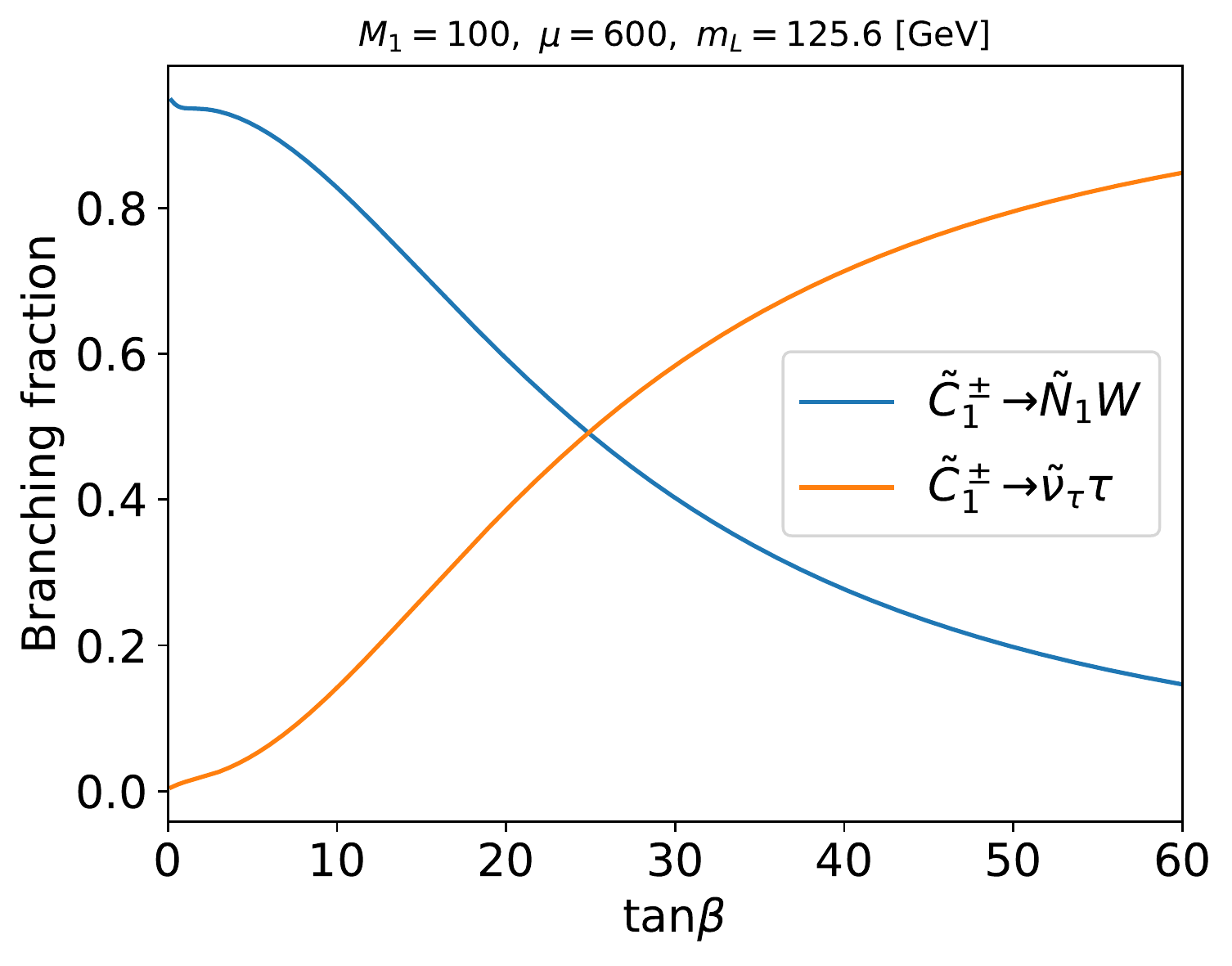}
  \end{minipage}         
  \end{tabular}
  \caption{Branching fractions of the higgsino-like neutralino and chargino.
    The top left (right) panel shows the branching fractions of $\tilde{N}_2$ ($\tilde{C}_1^{\pm}$)
    in the BHR scenario,
    while the bottom left (right) panel shows the branching fractions of $\tilde{N}_2$ ($\tilde{C}_1^{\pm}$)
    in the BHL scenario.
    Relevant parameters are shown in each figure,
    and others are taken to be larger than 3 TeV.
    }
  \label{fig:br-tanb}
\end{figure}
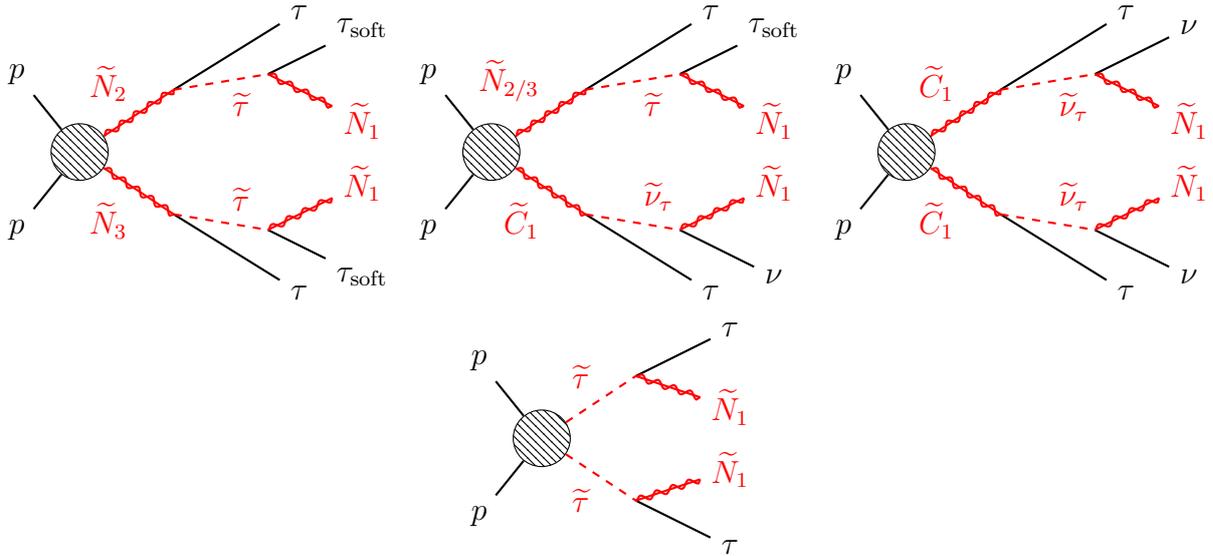
\begin{figure}[t]
  \centering
  \begin{minipage}{0.32\linewidth}
    \centering
    \begin{tikzpicture}
      \begin{feynman}
        \vertex (p1){\( p \)};
        \vertex[below =5em of p1] (p2){\( p \)};
        \vertex at ($(p1)!0.5!(p2) + (2em, 0)$) [blob, draw=black, pattern color=black] (v){};
        \vertex at ($(v) + (3em, 2em)$) (d1);
        \vertex at ($(d1) + (4em, 2.5em)$) (tau1){\(\tau\)};
        \vertex at ($(d1) + (3em, 0.5em)$) (d2);
        \vertex at ($(d2) + (3em, 1.5em)$) (st1){\(\tau_{\mathrm{soft}}\)};
        \vertex at ($(d2) + (3em, -1.5em)$) (n1){\(\textcolor{red}{\tilde{N}_1}\)};
        \vertex at ($(v) + (3em, -2em)$) (e1);
        \vertex at ($(e1) + (4em, -2.5em)$) (tau2){\(\tau\)};
        \vertex at ($(e1) + (3em, -0.5em)$) (e2);
        \vertex at ($(e2) + (3em, -1.5em)$) (st2){\(\tau_{\mathrm{soft}}\)};
        \vertex at ($(e2) + (3em, 1.5em)$) (n2){\(\textcolor{red}{\tilde{N}_1}\)};
          \diagram*{
            (p1) -- (v) -- (p2),
            (v) --[plain, boson, edge label=\(\tilde{N}_2\), red] (d1) -- (tau1),
            (d1) --[scalar, edge label'=\(\tilde{\tau}\), red] (d2) -- (st1),
            (d2) --[plain, boson, red] (n1),
            (v) --[plain, boson, edge label'=\(\tilde{N}_3\), red] (e1) -- (tau2),
            (e1) --[scalar, edge label=\(\tilde{\tau}\), red] (e2) -- (st2),
            (e2) --[plain, boson, red] (n2),
            };
      \end{feynman}
    \end{tikzpicture}
  \end{minipage}
  \begin{minipage}{0.32\linewidth}
    \centering
    \begin{tikzpicture}
      \begin{feynman}
        \vertex (p1){\( p \)};
        \vertex[below =5em of p1] (p2){\( p \)};
        \vertex at ($(p1)!0.5!(p2) + (2em, 0)$) [blob, draw=black, pattern color=black] (v){};
        \vertex at ($(v) + (3em, 2em)$) (d1);
        \vertex at ($(d1) + (4em, 2.5em)$) (tau1){\(\tau\)};
        \vertex at ($(d1) + (3em, 0.5em)$) (d2);
        \vertex at ($(d2) + (3em, 1.5em)$) (st1){\(\tau_{\mathrm{soft}}\)};
        \vertex at ($(d2) + (3em, -1.5em)$) (n1){\(\textcolor{red}{\tilde{N}_1}\)};
        \vertex at ($(v) + (3em, -2em)$) (e1);
        \vertex at ($(e1) + (4em, -2.5em)$) (tau2){\(\tau\)};
        \vertex at ($(e1) + (3em, -0.5em)$) (e2);
        \vertex at ($(e2) + (3em, -1.5em)$) (st2){\(\nu\)};
        \vertex at ($(e2) + (3em, 1.5em)$) (n2){\(\textcolor{red}{\tilde{N}_1}\)};
          \diagram*{
            (p1) -- (v) -- (p2),
            (v) --[plain, boson, edge label=\(\tilde{N}_{2/3}\), red] (d1) -- (tau1),
            (d1) --[scalar, edge label'=\(\tilde{\tau}\), red] (d2) -- (st1),
            (d2) --[plain, boson, red] (n1),
            (v) --[plain, boson, edge label'=\(\tilde{C}_1\), red] (e1) -- (tau2),
            (e1) --[scalar, edge label=\(\tilde{\nu}_{\tau}\), red] (e2) -- (st2),
            (e2) --[plain, boson, red] (n2),
            };
      \end{feynman}
    \end{tikzpicture}
  \end{minipage}
  \begin{minipage}{0.32\linewidth}
    \centering
    \begin{tikzpicture}
      \begin{feynman}
        \vertex (p1){\( p \)};
        \vertex[below =5em of p1] (p2){\( p \)};
        \vertex at ($(p1)!0.5!(p2) + (2em, 0)$) [blob, draw=black, pattern color=black] (v){};
        \vertex at ($(v) + (3em, 2em)$) (d1);
        \vertex at ($(d1) + (4em, 2.5em)$) (tau1){\(\tau\)};
        \vertex at ($(d1) + (3em, 0.5em)$) (d2);
        \vertex at ($(d2) + (3em, 1.5em)$) (st1){\(\nu\)};
        \vertex at ($(d2) + (3em, -1.5em)$) (n1){\(\textcolor{red}{\tilde{N}_1}\)};
        \vertex at ($(v) + (3em, -2em)$) (e1);
        \vertex at ($(e1) + (4em, -2.5em)$) (tau2){\(\tau\)};
        \vertex at ($(e1) + (3em, -0.5em)$) (e2);
        \vertex at ($(e2) + (3em, -1.5em)$) (st2){\(\nu\)};
        \vertex at ($(e2) + (3em, 1.5em)$) (n2){\(\textcolor{red}{\tilde{N}_1}\)};
          \diagram*{
            (p1) -- (v) -- (p2),
            (v) --[plain, boson, edge label=\(\tilde{C}_1\), red] (d1) -- (tau1),
            (d1) --[scalar, edge label'=\(\tilde{\nu}_{\tau}\), red] (d2) -- (st1),
            (d2) --[plain, boson, red] (n1),
            (v) --[plain, boson, edge label'=\(\tilde{C}_1\), red] (e1) -- (tau2),
            (e1) --[scalar, edge label=\(\tilde{\nu}_{\tau}\), red] (e2) -- (st2),
            (e2) --[plain, boson, red] (n2),
            };
      \end{feynman}
    \end{tikzpicture}
  \end{minipage}
    \begin{minipage}{1.0\linewidth}
    \centering
    \begin{tikzpicture}
      \begin{feynman}
        \vertex (p1){\( p \)};
        \vertex[below =5em of p1] (p2){\( p \)};
        \vertex at ($(p1)!0.5!(p2) + (2em, 0)$) [blob, draw=black, pattern color=black] (v){};
        \vertex at ($(v) + (3em, 2em)$) (d1);
        \vertex at ($(d1) + (3em, 1.5em)$) (tau1){\(\tau\)};
        \vertex at ($(d1) + (3em, -1.0em)$) (n1){\(\textcolor{red}{\tilde{N}_1}\)};
        \vertex at ($(v) + (3em, -2em)$) (e1);
        \vertex at ($(e1) + (3em, -1.5em)$) (tau2){\(\tau\)};
        \vertex at ($(e1) + (3em, 1.0em)$) (n2){\(\textcolor{red}{\tilde{N}_1}\)};
          \diagram*{
            (p1) -- (v) -- (p2),
            (v) --[scalar, edge label=\(\tilde{\tau}\), red] (d1) -- (tau1),
            (d1) --[plain, boson, red] (n1),
            (v) --[scalar, edge label'=\(\tilde{\tau}\), red] (e1) -- (tau2),
            (e1) --[plain, boson, red] (n2),
            };
      \end{feynman}
    \end{tikzpicture}
  \end{minipage}
  \caption{
  Top: The Feynman diagrams of the neutralino/chargino production that produce two hard taus in the present scenarios,
 corresponding to the processes \eqref{eq:lhc-bhr}, \eqref{eq:lhc-bhl-nc}, and \eqref{eq:lhc-bhl-cc}.
 Bottom: Direct production of staus decaying into taus and LSPs, which is the original target of the ATLAS search we recast~\cite{ATL-PHYS-PUB-2016-021}. In our scenarios, this process will not yield signal events because the taus from this process are soft.}
  \label{fig:diagram}
\end{figure}

On the other hand, the higgsino is not degenerate with the bino LSP,
and is expected to provide  viable signals at the HL-LHC.
Figure~\ref{fig:br-tanb} shows the branching fractions of $\tilde{N}_2$ and $\tilde{C}_1^{\pm}$ calculated with \texttt{SUSY-HIT} \cite{Djouadi:2006bz} as functions of $\tan\beta$ at typical model points:
$(M_1, \mu, m_R)=(100, -600, 108)\GeV$ for the BHR scenario, and
$(M_1, \mu, m_L)=(100, 600, 125.6)\GeV$ for the BHL scenario.
These parameters provide the correct DM relic abundance,
and $g-2$ discrepancy becomes less than $2\sigma$ for $\tan\beta=40$
(see Fig.~\ref{fig:result}).
Since the higgsino decays into a tau and a stau are governed by the tau Yukawa coupling $y_{\tau}=m_{\tau}/v\cos\beta$,
its partial width is enhanced for large $\tan\beta$. We see that, for $\tan\beta=40$, its branching ratio is as large as
$\mathrm{Br}(\tilde{N}_2\to \tilde{\tau}\tau)
\sim
\mathrm{Br}(\tilde{C}_1^{\pm}\to \tilde{\tau}\nu_{\tau})
\sim
0.7$.
We checked that the branching ratio of $\tilde{N}_3$ has the similar behavior.
\footnote{The branching fraction of $\tilde{N}_3\to \tilde{\tau}\tau$ is similar to that of $\tilde{N}_2\to \tilde{\tau}\tau$, while the branching fraction of $\tilde{N}_3 \to Z (h)$ has similar behavior to that of $\tilde{N}_2 \to h (Z)$.}
Changing $M_1, \mu, m_R$, or $m_L$ for fixed $\tan\beta$ does not change the situation much.
Therefore, a large fraction of the events with chargino/neutralino pair production leads to two hard taus and missing transverse energy (cf.\ the top panels of Fig.~\ref{fig:diagram}):
\begin{align}
& pp\to\tilde{N}_2\;  \tilde{N}_3 \to  
\tau \tilde{\tau}\;  \tau \tilde{\tau}
\to
\tau \tau_{\rm soft}\tilde{N}_1 
\;
  \tau \tau_{\rm soft}\tilde{N}_1
  \quad \text{(BHR and BHL)},
  \label{eq:lhc-bhr}
\\
& pp\to\tilde{N}_{2/3}\;  \tilde{C}_1 \to  
\tau \tilde{\tau}\;  \tau \tilde{\nu}
\to
\tau \tau_{\rm soft}\tilde{N}_1 
\;
\tau \nu\tilde{N}_1 
  \quad \text{(BHL)},
  \label{eq:lhc-bhl-nc}
\\
& pp\to\tilde{C}_1\;  \tilde{C}_1 \to  
\tau \tilde{\nu}\;  \tau \tilde{\nu}
\to
\tau \nu\tilde{N}_1 
\;
\tau \nu\tilde{N}_1 
  \quad \text{(BHL)}.
  \label{eq:lhc-bhl-cc}
\end{align}
Thus, searches for events with two hadronic taus and missing transverse momentum, which are originally designed to search for the stau pair production ($pp\to \tilde{\tau}\tilde{\tau}\to \tau \tilde{N}_1 \tau \tilde{N}_1$), can probe the current scenarios.\footnote{We have also considered the prospects of the searches for $WZ$ and $Wh$ channel using an ATLAS study~\cite{ATL-PHYS-PUB-2014-010}, but found that they are much weaker than searches for two taus and missing energy.}
In the next subsection, we investigate the future prospect of searches for two taus and missing energy at the HL-LHC, recasting the result by the ATLAS \cite{ATL-PHYS-PUB-2016-021}.

Let us briefly discuss the current constraint from the $13\TeV$ LHC on the BHR and BHL scenarios, based on the recent ATLAS result on a search for neutralino/chargino production in events with at least two hadronic taus~\cite{ATLAS-CONF-2016-093}.
They analyzed the simplified model that consists of $\tilde{N}_2$ and $\tilde{C}_1^\pm$ being wino-like,
$\tilde{N}_1$ as the bino-like LSP, and left-handed stau/tau-sneutrino $\tilde{\tau}_L, \tilde{\nu}_\tau$.
The other sparticles are assumed to be heavy.
The masses of $\tilde{\tau}_L$ and $\tilde{\nu}_\tau$ are set to be halfway between the masses of $\tilde{C}_1^\pm$ and $\tilde{N}_1$.
They obtained an upper limit on wino-like chargino mass of $580\TO520\GeV$ for the LSP mass of $0\TO150\GeV$,
and no limit is obtained for $m_{\tilde N_1}>150\GeV$.
Simply comparing the cross section of wino-like chargino to that of the processes \eqref{eq:lhc-bhr}--\eqref{eq:lhc-bhl-cc},
we may have a rough estimate on the upper limit on higgsino mass; for $M_1=100\GeV$, it would be $-\mu=302\GeV$ for the BHR scenario and
$\mu=490\GeV$ for the BHL scenario,
which do not exceed the XENON1T sensitivity.
Since models with $m_{\tilde{N}_1}>150\GeV$ are not constrained in this results, we expect that the constraint from the $13\TeV$ LHC is limited to exclude a part of the $1\sigma$ parameter space.
Therefore, there still remains a wide region of the parameter space avoiding current constraints and the XENON1T prospect in the respective models, and it is important to study HL-LHC prospect to cover the whole parameter space.

\subsubsection{HL-LHC Prospect}
To evaluate the sensitivity of the HL-LHC to our scenarios,
we recast the result of the ATLAS study in Ref.~\cite{ATL-PHYS-PUB-2016-021},
which investigated the future sensitivity of a search for the direct stau production in events with at least two hadronic taus and missing transverse momentum at the HL-LHC (cf.\ the bottom panel of Fig.~\ref{fig:diagram}).
The ATLAS event selection requires at least one opposite-sign tau pair decaying to hadrons.
They considered three simplified models: left-handed stau production ($pp\to\tilde\tau_L\tilde \tau_L$), right-handed stau production ($pp\to\tilde\tau_R\tilde \tau_R$), and both of the left- and right-handed stau production ($pp\to\tilde\tau_{L,R}\tilde\tau_{L,R}$), in which the staus decay into a tau and the LSP.
For instance, in the case of combined $\tilde{\tau}_L+\tilde{\tau}_R$ production
with the 30\% systematic uncertainty on the SM background,
the upper bound is $m_{\tilde\tau}=710\GeV$ for $m_{\tilde{N}_1}=100\GeV$,
which corresponds to the stau pair production cross section of 0.14 fb,
and no limit is obtained for $m_{\tilde{N}_1} \gtrsim \SI{300}{GeV}$.
For pure $\tilde{\tau}_L\tilde{\tau}_L$ and pure $\tilde{\tau}_R\tilde{\tau}_R$ production,
the situations are similar and we can obtain the upper bound on production cross section of $0.14\TO0.15\fb$
if the mass difference between the stau and the LSP is large enough.

\begin{figure}[t]
  \centering
    \includegraphics[width=0.6\textwidth]{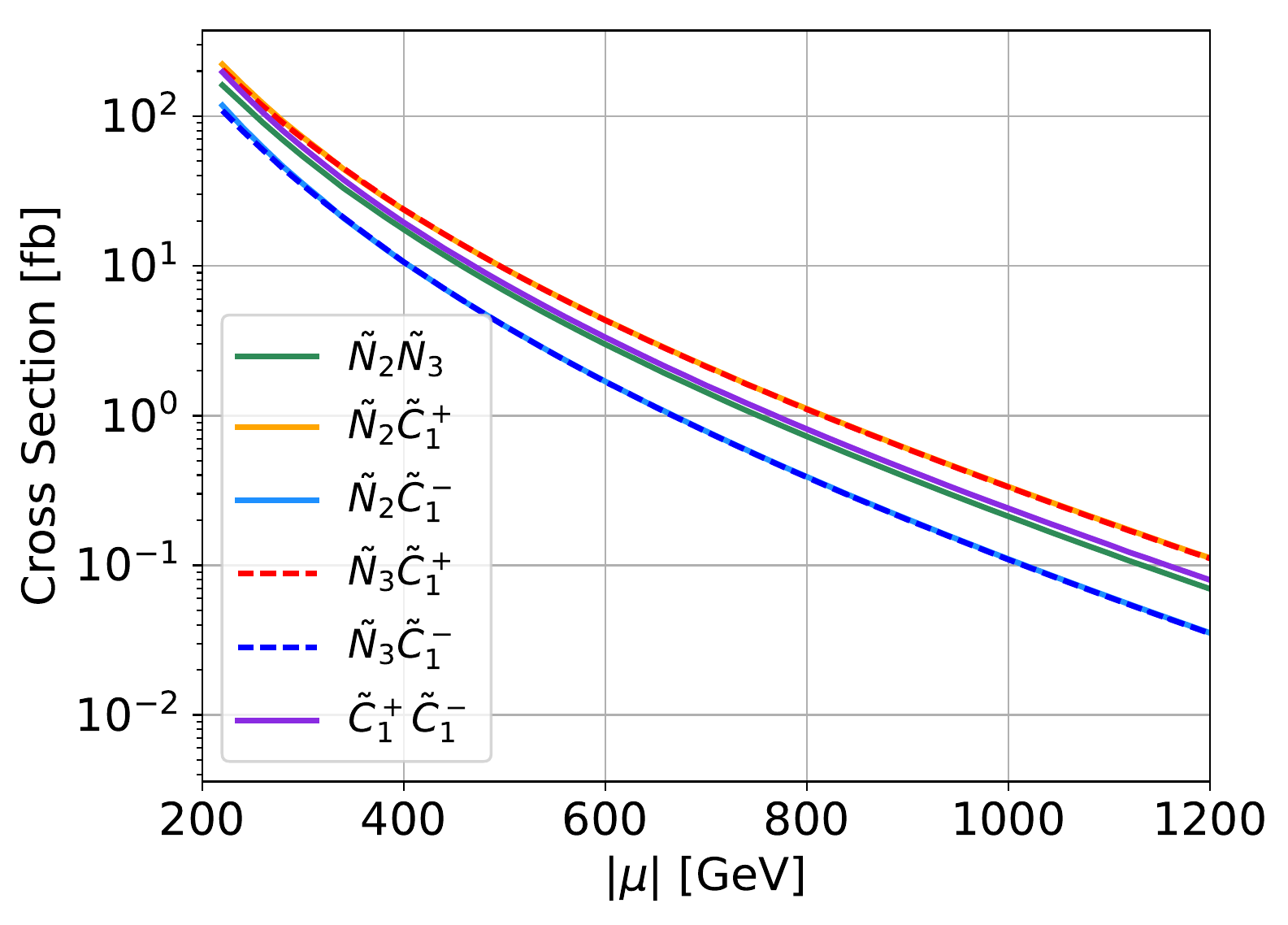}
  \caption{The production cross section of higgsino-like neutralino/chargino at the 14 TeV LHC
    with $M_1=\SI{100}{GeV}$ and $M_2=\SI{3}{TeV}$.}
  \label{fig:xsect-higgsino}
\end{figure}

In order to recast these results,
we calculate the NLO production cross section of neutralino/chargino by \texttt{Prospino 2}~\cite{Beenakker:1996ed}, which is shown in Fig.~\ref{fig:xsect-higgsino}, and
their branching fractions to the modes involving a hard tau by \texttt{SUSY-HIT}~\cite{Djouadi:2006bz}, shown in Fig.~\ref{fig:br-tanb}.
We consider the process \eqref{eq:lhc-bhr} for the BHR scenario, while
the processes \eqref{eq:lhc-bhr}--\eqref{eq:lhc-bhl-cc} are taken into account for the BHL scenario.
Ignoring the soft particles $\tau_{\mathrm{soft}}$ and $\nu$, we compare the effective cross sections of these processes to that of combined $\tilde{\tau}_L+\tilde{\tau}_R$ production in the ATLAS study.
Note that the signal acceptances for the processes \eqref{eq:lhc-bhr} and \eqref{eq:lhc-bhl-nc} decrease by a factor $1/2$ because the two hard taus do not necessarily have opposite sign.
As is done in the ATLAS analysis, we consider three different systematic uncertainties of 20, 30, and 50\%.

In Fig.~\ref{fig:result}, the green dashed lines show the 95\% CL upper limits on $|\mu|$ in the presence of 30\% systematic uncertainty,
and hatched regions correspond to  20--50\% systematic uncertainty.
We can see that 95\% CL upper limit is $-\mu=800\TO850\GeV$ for the BHR scenario
and $\mu=1140\TO1200\GeV$ for the BHL scenario,
and all the parameter space explaining the muon $g-2$ discrepancy will be covered by this analysis.
Note that we are interested in $M_1 \lesssim \SI{200}{GeV}$,
which can explain the muon $g-2$ discrepancy and avoids the LUX constraint,
and hence the mass difference $|\mu| - M_1$ is large enough to have a similar acceptance as the models ATLAS analyzed, which allows us to estimate the constraint simply by comparing the cross sections.

\subsection{ILC prospects}
\label{sec:ilc-prospects}
As a further experimental research in the future, we consider the ILC
with $\sqrt s=500\GeV$ (ILC500).
At the ILC, slepton--anti-slepton pairs are produced from $e^+e^-$ collision via the $Z$-boson and photon exchange.
With its clean environment, the ILC can probe productions of NLSP sleptons that are degenerate with the LSP, contrary to the LHC.
The ILC500 can exclude, for instance, the NLSP $\tilde{\mu}_R$ up to 248 GeV at 95\% CL for the mass difference smaller than 10 GeV~\cite{Baer:2013vqa}.
This prospect is applied to our scenario as described in Fig.~\ref{fig:result} by the red solid lines; the left region to the lines can be probed at the ILC.

\subsection{Impact of the wino mass}
\label{subsec:M2}

\begin{figure}
  \centering
  \includegraphics[width=0.5\textwidth]{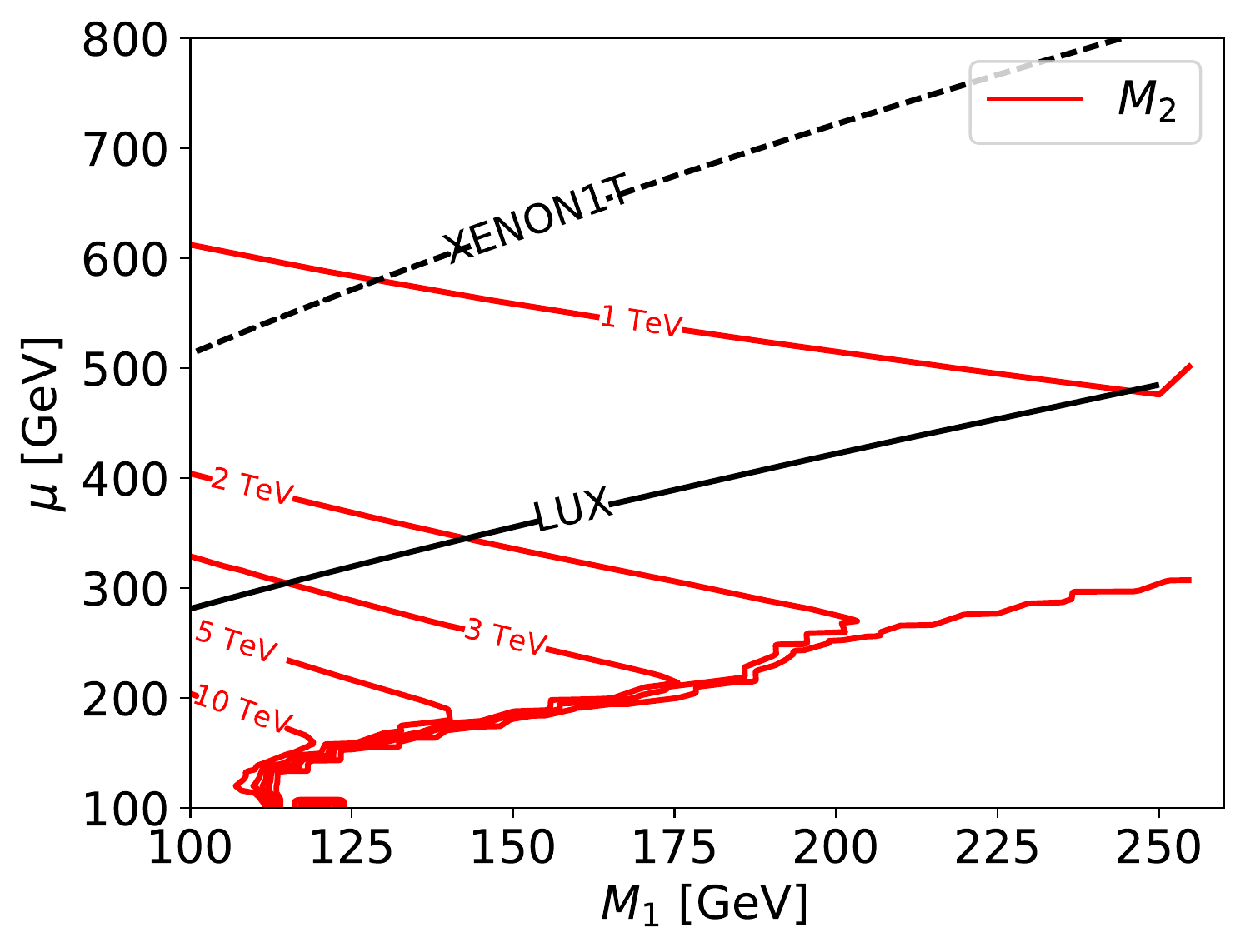}
 \caption{The upper bound on $M_2$ that can explain the muon $g-2$ discrepancy at the $1\sigma$ level in the BHL scenario.
  The other parameters are the same of the right panel of Fig.~\ref{fig:result}.}
  \label{fig:wino-contour}
\end{figure}

As shown in Fig.~\ref{fig:result}, the BHL scenario requires a smaller value of the higgsino mass than the BHR case to explain the $g-2$ discrepancy, which is due to the smaller factor in Eq.~\eqref{eq:BHL} than in Eq.~\eqref{eq:BHR}.
On the other hand, the BHL scenario has broader parameter space for the muon $g-2$
in the large $M_1$ region.
This is because we assume $M_2 = \SI{3}{TeV}$ as a heavy wino mass.
Equations~\eqref{eq:WHL1} and \eqref{eq:WHL2} tell us
that the WHL contributions are large due to the SU(2) gauge coupling.
In fact, $a_{\mu}(\mathrm{BHL})\sim a_{\mu}(\mathrm{WHL1}) +a_{\mu}(\mathrm{WHL2})$ is realized in most of the parameter space in Fig.~\ref{fig:result}.
To study the impact of the wino mass in the BHL scenario,
we show in Fig.~\ref{fig:wino-contour} the contours of upper bounds on $M_2$ to explain the muon $g-2$ discrepancy at the $1\sigma$ level.
DM abundance and direct detection constraints are the same as the right panel of Fig.~\ref{fig:result} as far as $M_2\gg \mu$.
On the other hand, for $M_2\simeq\mu$,
the HL-LHC prospect is altered due to the non-negligible wino component
in neutralino/chargino.
For instance, if $M_2 = 1\TeV$, the higgsino-wino mixing decreases the branching fraction of neutralino/chargino
into $\tilde{\tau}\tau$/$\tilde{\nu}_\tau\tau$,
which weakens the exclusion reach at HL-LHC to $935 \TO 960 \GeV$
depending on the systematic uncertainty.
The HL-LHC cannot cover the whole parameter space to explain the muon $g-2$ discrepancy at $2\sigma$.
In this case, in addition to the stau-search at the HL-LHC,
we need the XENON1T and the ILC500 in order to cover most of the parameter space.

\section{Summary and discussion}
\label{sec:summary-discussion}
We have shown that bino-higgsino-slepton scenarios can explain the muon $g-2$ discrepancy and the DM relic abundance simultaneously.
Much of the parameter space will be probed by the XENON1T,
while at the HL-LHC the whole parameter space will be probed by searches for events with two hadronic taus and missing transverse momentum.
The ILC500 can further test the scenarios through the direct production of sleptons.

We have taken $\tan\beta=40$ throughout our analysis.
If we take a larger value for $\tan\beta$,
the preferred parameter space becomes broader,
while the sensitivity of the HL-LHC search is also strengthened because the higgsino branching fractions to stau/tau-sneutrino get closer to unity.

The assumption of the universal slepton mass can also be relaxed,
as long as smuon is light enough to produce sufficient contributions to the muon $g-2$
and at least one of the slepton masses is close to the LSP mass in order to provide the correct DM abundance.
The HL-LHC search is still prospective in this case provided that the stau is sufficiently lighter than the higgsinos.

In the parameter regions with $M_1<\SI{100}{GeV}$, which we did not consider in the analyses,  the DM abundance and the SUSY contribution to the muon $g-2$ have nontrivial dependences on the higgsino and slepton masses.
In particular, the LSP annihilation cross section is enhanced
at $m_{\tilde{N}_1} \sim m_Z/2$ or $m_h/2$.
These possibilities are left to be studied in future works.

Let us briefly mention other SUSY scenarios to explain the muon $g-2$ discrepancy.
The BLR scenario, where only the bino and the left- and right-handed sleptons are light,
can also provide a minimal explanation of the muon $g-2$ and DM.
As shown in Eq.~\eqref{eq:BLR}, the contribution to the muon $g-2$ is proportional to $\mu\tan\beta$.
Increasing $\mu\tan\beta$ provides a sizable contribution to the muon $g-2$,
and the bino-like neutralino can be the DM in presence of the coannihilation effect with the sleptons.
For this scenario, we should also be aware of the constraint from the vacuum stability as well as from the DM direct detection and collider experiments,
since large $\mu\tan\beta$ causes the charge-breaking minima in the potential
(see, e.g., Ref.~\cite{Endo:2013lva}).

Allowing more than three SUSY multiplets having $\Order(100)\GeV$ masses, we can go beyond the minimal scenarios we have considered.
Here, the wino mass plays a crucial role.
As seen in Fig.~\ref{fig:wino-contour}, wino contribution to muon $g-2$ is sizable even for $M_2>\SI{1}{TeV}$, provided higgsino and left-handed sleptons are relatively light.
Lighter wino will allow us to solve the muon $g-2$ discrepancy in a larger parameter space, and also makes the HL-LHC prospects more involved; for example, the stau search becomes less effective as discussed in Sec.~\ref{subsec:M2}.

In this letter, we have studied constraints and future prospects of SUSY models in which the muon $g-2$ discrepancy and the dark matter relic abundance are simultaneously explained. In the near future, the sensitivity of the dark matter direct detection experiment will be improved by XENON1T and other experiments, 
and the new experiments at Fermilab~\cite{Grange:2015fou,Chapelain:2017syu} and J-PARC~\cite{Iinuma:2011zz} will provide more precise measurements on the muon $g-2$.
Furthermore, the lattice QCD calculations of the SM hadronic light-by-light contributions are now in progress (see, e.g., \cite{Blum:2016lnc}), and hence we expect a significant progress on the muon $g-2$ measurement in both experimental and theoretical sides. 
We hope this letter is useful for further studies towards the searches for new physics in the light of muon $g-2$ and DM.

\section*{Acknowledgment}
We thank Takahiro Yoshinaga for the collaboration in the early stage of this work.
We also thank
Sascha Caron, 
Monica D'Onofrio, 
Tina Potter, 
Krzysztof Rolbiecki 
and Koji Terashi 
for useful discussion.
This work is supported by the Grant-in-Aid for Scientific Research (No.~16K17681 [ME], No.~16H03991 [ME], No.~26104001 [KH], No~.26104009 [KH], No.~26247038 [KH], No.~26800123 [KH], No.~16H02189 [KH]), and by World Premier International Research Center Initiative (WPI Initiative), MEXT, Japan.
S.I. is supported in part at the Technion by a fellowship from the Lady Davis Foundation, and also by the Israel Science Foundation (Grant No.~720/15), by the United-States-Israel Binational Science Foundation (BSF) (Grant No.~2014397), and by the ICORE Program of the Israel Planning and Budgeting Committee (Grant No.~1937/12).
Diagrams in Fig.~\ref{fig:diagram} are drawn with \texttt{TikZ-Feynman}~\cite{Ellis:2016jkw}.

\bibliographystyle{utphys27mod}
\bibliography{ref}

\end{document}
